\documentclass[conference]{IEEEtran}

\usepackage{cite}
\usepackage{amsmath,amssymb,amsfonts}
\usepackage{algorithmic}
\usepackage{graphicx}
\usepackage{textcomp}
\usepackage{xcolor}
\usepackage{hyperref}
\usepackage[ruled,vlined]{algorithm2e} 

\begin{document}

\title{Optimizing High-Throughput Distributed Data Pipelines for Reproducible Deep Learning at Scale}

\author{
\IEEEauthorblockN{Kashish Mittal, Di Yu, Roozbeh Ketabi, Arushi Arora, Brendon Lapp, and Peng Zhang}
\IEEEauthorblockA{\textit{AI Platform \& Matching ML} \\
\textit{Uber Technologies}\\
San Francisco, CA, USA \\
\{kashish, yud, roozbeh, arushi.arora, getlapped, pengz\}@uber.com}
}

\maketitle

\begin{abstract}
Training massive-scale deep learning models on datasets spanning tens of terabytes presents critical challenges in hardware utilization and training reproducibility. In this paper, we identify and resolve profound data-loading bottlenecks within distributed GPU training pipelines using the Petastorm data loader and Apache Parquet datasets. Through systematic profiling, we demonstrate that network I/O and CPU-bound data transformations (e.g., PyArrow to NumPy) constrain GPU utilization to as low as 10-15\%. To address this, we propose an optimized architecture that features push-down worker-level transformations coupled with local-disk caching via FanoutCache, minimizing redundant I/O and CPU overhead across training epochs. Furthermore, we eliminate race conditions in multi-worker shared queues by implementing dedicated round-robin ventilator and result queues, alongside modernized RNG handling, achieving strict deterministic data loading. Our optimizations yield a 6$\times$ speedup, reducing end-to-end training time from 22 hours to 3 hours, increasing GPU utilization to over 60\%, and drastically reducing run-to-run variance, enabling robust, high-throughput, and reproducible large-scale model training.
\end{abstract}


\section{Introduction}
The scale of industrial machine learning has grown exponentially, heavily relying on the processing of massive tabular datasets spanning tens of terabytes, tens of billions of rows, and hundreds of features. At this scale, infrastructure costs and training times become paramount. However, as GPU compute capabilities rapidly advance, data loading pipelines frequently fail to keep pace. This divergence leads to critical I/O bottlenecks and GPU starvation; in our baseline industrial deep learning pipelines, we observed high-performance GPUs idling at 10--15\% utilization while waiting for data batches. 

Compounding this throughput issue is the challenge of reproducibility. In highly concurrent, multi-worker distributed training setups, identical configurations and fixed seeds frequently yield significant run-to-run variance in key evaluation metrics. This stochasticity stems from hidden race conditions in data delivery, making it exceedingly difficult to reliably compare novel model architectures against production baselines. 

In this work, we present an optimized architecture for the Petastorm data loader: a library for reading Apache Parquet\texttrademark{} datasets, that resolves both the throughput and determinism bottlenecks. Our primary contributions are:
\begin{itemize}
    \item \textbf{Push-down Data Transformations:} Offloading CPU-heavy data formatting (e.g., PyArrow-to-NumPy conversions) directly to the worker pool to unblock the main training thread.
    \item \textbf{Quota-Managed Local Disk Caching:} Utilizing \texttt{FanoutCache} to eliminate redundant network I/O across training epochs for datasets that exceed local node storage capacities.
    \item \textbf{Deterministic Worker Scheduling:} Replacing shared queues with dedicated, round-robin ventilator and result queues to eliminate multi-threading race conditions and guarantee strict data reproducibility.
\end{itemize}

Through empirical evaluation on production-scale deep learning models, our proposed optimizations yield a 6$\times$ overall speedup, reducing end-to-end training time from 22 hours to approximately 3 hours. By saturating the training pipeline, GPU utilization is increased to over 60\%, slashing compute costs by nearly 80\%. Furthermore, our scheduling redesign reduces run-to-run mean average precision (MAP) shift to negligible levels, enabling robust and reproducible model development at scale.

\section{Background \& Related Work}

\subsection{Distributed Training Infrastructure}
The foundation of our large-scale deep learning stack relies on three primary components: Ray, Horovod, and Petastorm. 
Training jobs are orchestrated using Ray clusters, which provide flexible and distributed resource management across worker nodes. For multi-GPU synchronization, we utilize Horovod, which leverages ring-allreduce algorithms to perform efficient distributed gradient descent. 

Crucially, our datasets are stored as Apache Parquet\texttrademark{} files on the Hadoop Distributed File System (HDFS). Because native deep learning frameworks lack built-in capabilities for efficiently streaming large-scale, distributed Parquet data, we utilize Petastorm, an open-source data access library designed specifically to bridge the gap between HDFS storage and deep learning frameworks (e.g., TensorFlow and PyTorch). Petastorm facilitates the reading of specific row groups, transforming columnar data into dense tensors required for neural network ingestion.

\subsection{Related Work}
Data loading bottlenecks in deep learning are a widely recognized challenge, and several solutions exist to address them. Standard framework utilities, such as PyTorch's \texttt{DataLoader} and TensorFlow's \texttt{tf.data}, offer robust pipeline abstractions. However, when dealing with out-of-core tabular data stored in Parquet format across a network, these native loaders often require custom, unoptimized parsers that fail to scale efficiently.

Alternatively, hardware-accelerated pipelines like NVIDIA DALI have become the industry standard for eliminating CPU bottlenecks by offloading data decoding and augmentation directly to the GPU. While highly effective for computer vision tasks (e.g., JPEG decoding and image resizing), DALI is not natively optimized for the massive serialization and deserialization overhead inherent to distributed tabular data parsing, where the CPU must handle complex PyArrow-to-NumPy transformations.

Other approaches involve utilizing distributed in-memory file systems (e.g., Alluxio or Redis) to bypass slow network I/O. However, loading tens of terabytes of training data entirely into RAM is cost-prohibitive and practically infeasible for our scale. Consequently, our work focuses on optimizing the existing Parquet-based pipeline to work within the strict memory constraints of local worker node disks.

\section{Challenge I: Alleviating Data Loading Bottlenecks}

\subsection{Profiling and Bottleneck Analysis}
When initial observations revealed GPU utilization hovering at an unacceptably low 10--15\%, we systematically isolated the root causes. First, to rule out inadequate model complexity, we ran synthetic benchmarks with significantly widened and deepened model architectures. Under these conditions, utilization jumped to $\sim$80\%, confirming the GPUs were effectively starving for data rather than lacking compute-bound work. 

\begin{figure}[htbp]
\centerline{\includegraphics[width=\columnwidth]{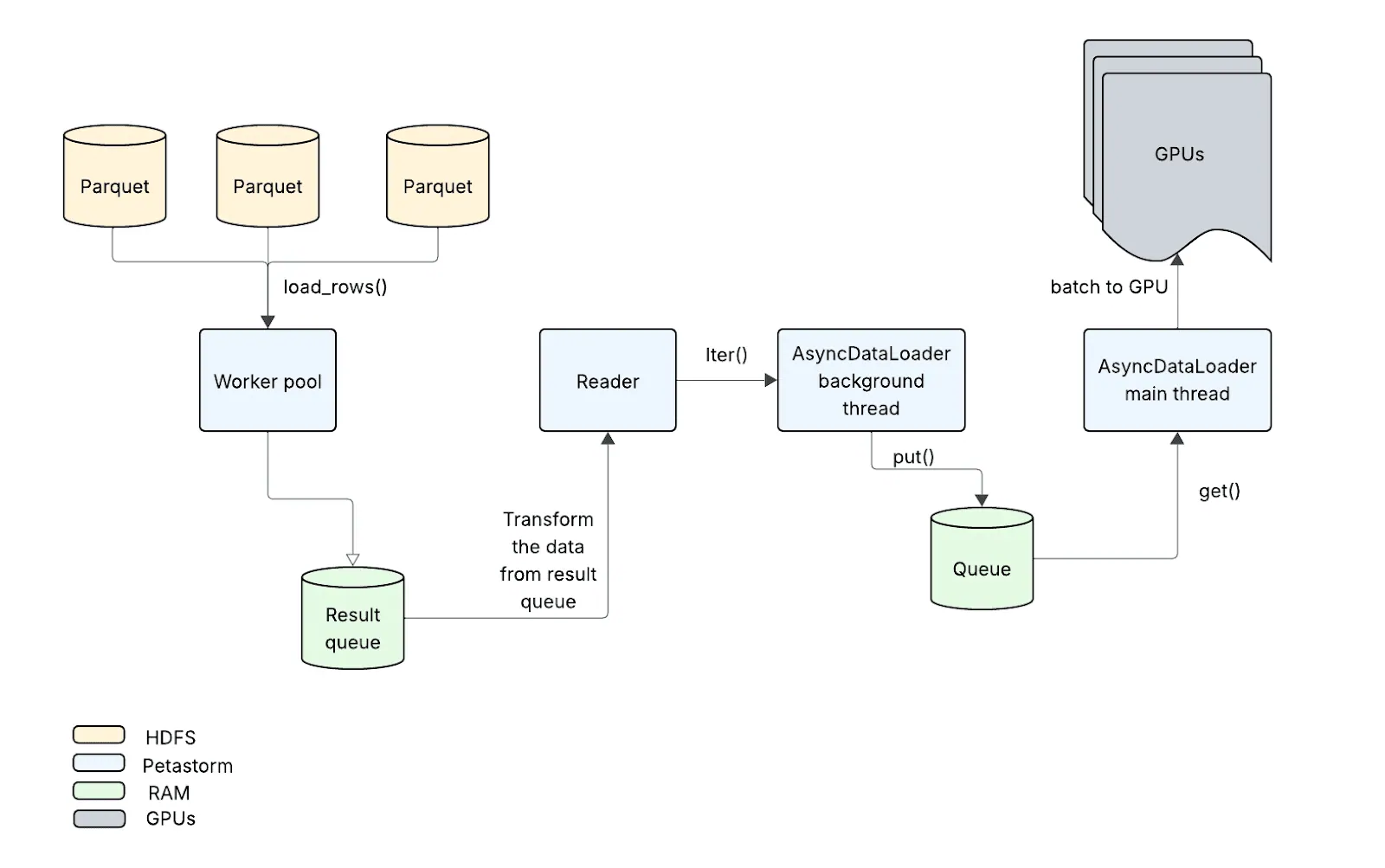}}
\caption{Baseline distributed training data flow. The main thread acts as a bottleneck by applying CPU-intensive transformations just-in-time.}
\label{fig:baseline_arch}
\end{figure}

Next, we evaluated network latency by forcing a subset of the dataset into an in-memory RAM cache, which spiked utilization to over 60\%, identifying HDFS network I/O as a primary bottleneck. However, a subsequent attempt to implement a raw local disk cache failed to improve utilization. This failure revealed a secondary, hidden bottleneck: CPU contention. While the local disk solved the network latency, the training pipeline was still performing synchronous, CPU-intensive data transformations (PyArrow to NumPy) just-in-time within the main thread, blocking batches from reaching the GPU (see Figure~\ref{fig:baseline_arch}). 

\subsection{Proposed Architecture}
To dismantle this dual I/O and CPU bottleneck, we architected a solution built on two pillars: push-down worker-level transformations and localized caching.

\subsubsection{Push-Down Transformations}
In the original architecture, the worker pool returned raw Apache Arrow tables. We modified the pipeline to push the transformation logic directly down to the Petastorm worker threads. Now, workers apply the PyArrow-to-NumPy conversion immediately after reading a row group. Consequently, the queue connecting the workers to the GPU is populated with pre-transformed, ready-to-train NumPy arrays. While this trades memory efficiency for throughput, it successfully parallelizes the CPU load and frees the main thread to focus exclusively on batch propagation.

\begin{figure}[htbp]
\centerline{\includegraphics[width=\columnwidth]{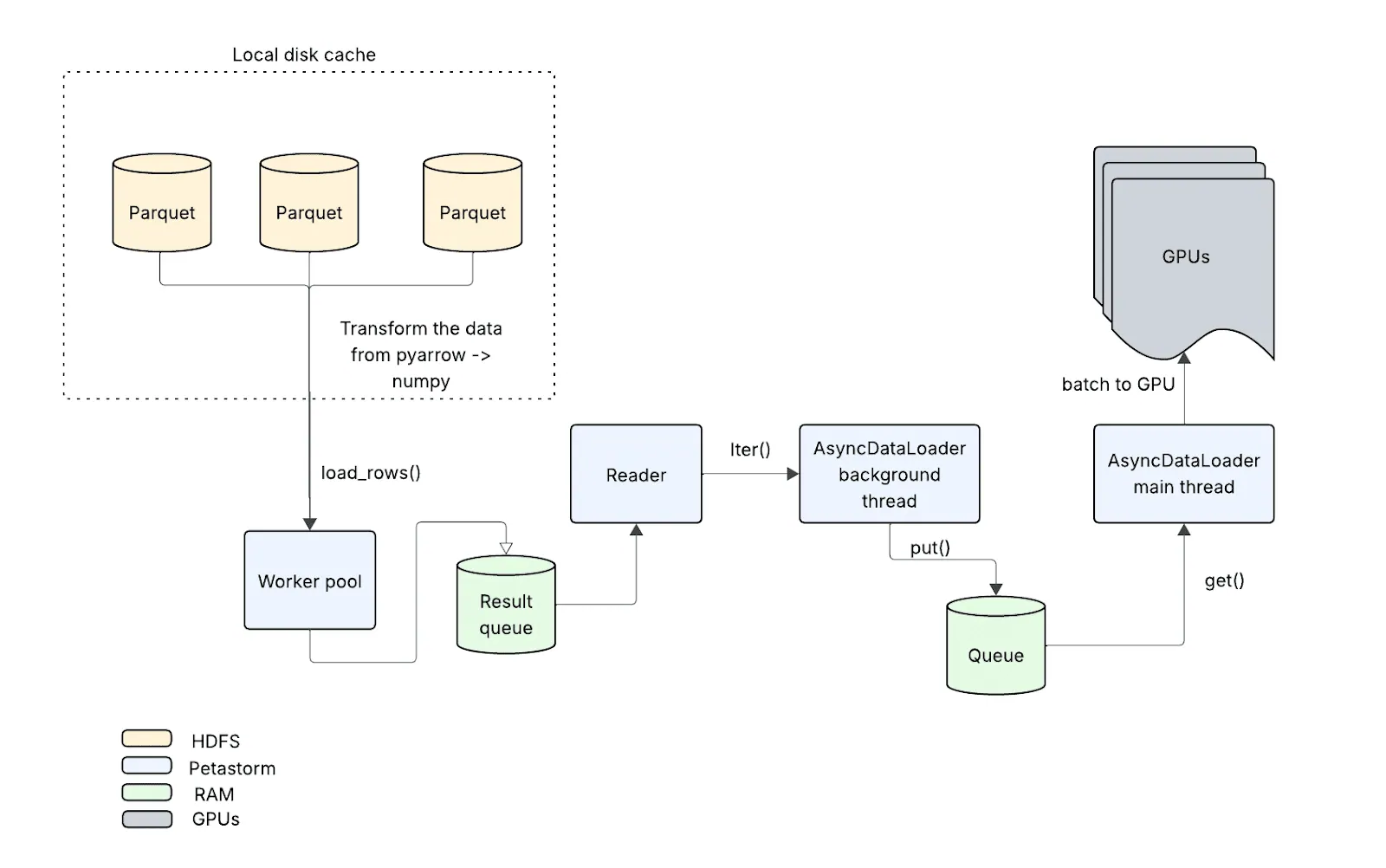}}
\caption{Optimized architecture leveraging push-down transformations to the worker pool and local disk caching via \texttt{FanoutCache}.}
\label{fig:optimized_arch}
\end{figure}

\subsubsection{Quota-Managed Local Disk Caching}
Because training occurs over multiple epochs, fetching the same row groups from HDFS repeatedly is highly inefficient. We implemented a local disk cache using the \texttt{diskcache.FanoutCache} library. Crucially, because datasets often exceed the storage capacity of a single worker node, standard eviction policies like Least Recently Used (LRU) are counterproductive for sequential epoch training. Instead, we implemented a quota management strategy: row groups are cached until the local disk quota is reached, after which the system falls back to HDFS reads. 

The algorithmic flow is detailed in Algorithm \ref{alg:worker_fetch}. By caching the \textit{pre-transformed} data, a cache hit entirely bypasses both network I/O and CPU transformation overhead (Figure~\ref{fig:optimized_arch}).

\begin{algorithm}[htbp]
\caption{Optimized Worker Pool Fetching}
\label{alg:worker_fetch}
\begin{algorithmic}[1]
\REQUIRE Row Group ID $R$, Local Disk Cache $C$, HDFS $H$
\IF{$R \in C$}
    \STATE $Data \leftarrow C.read(R)$ \COMMENT{Fast path: pre-transformed}
\ELSE
    \STATE $RawData \leftarrow H.read(R)$
    \STATE $Data \leftarrow \text{TransformToNumPy}(RawData)$
    \IF{$C.size < \text{Quota}$}
        \STATE $C.write(R, Data)$
    \ENDIF
\ENDIF
\RETURN $Data$
\end{algorithmic}
\end{algorithm}

\subsubsection{Thread Pool Stability}
To further optimize performance, we migrated from a Process Pool to a Thread Pool, eliminating the CPU overhead associated with pickling data for inter-process communication. To mitigate the risk of "zombie threads" hanging on slow I/O calls, we introduced robust sentinel logic in the ventilator to gracefully signal thread termination and tightened HDFS timeout policies.

\section{Challenge II: Ensuring Reproducibility in Distributed Contexts}

\subsection{Sources of Stochasticity}
In large-scale distributed training, guaranteeing deterministic data loading is essential for reliably evaluating and comparing novel model architectures against production baselines. While standard Petastorm explicitly controls high-level operations such as deterministic row-group shuffling and data sharding across GPU workers, we observed significant run-to-run variance in offline evaluation metrics despite using fixed seeds.

Systematic profiling revealed that this stochasticity did not originate from the shuffling algorithms, but rather from race conditions inherent to the multi-threading architecture. In the baseline system, both the ventilator queue (which distributes row groups to workers) and the result queue (which collects processed batches) were shared across all worker threads. Consequently, the exact order in which data was fetched and batched was dictated by unpredictable variables: OS thread scheduling, network I/O timing, and individual worker execution speeds. Paradoxically, as we scaled the number of workers to improve throughput, these race conditions became increasingly pronounced.

\begin{figure}[htbp]
\centerline{\includegraphics[width=\columnwidth]{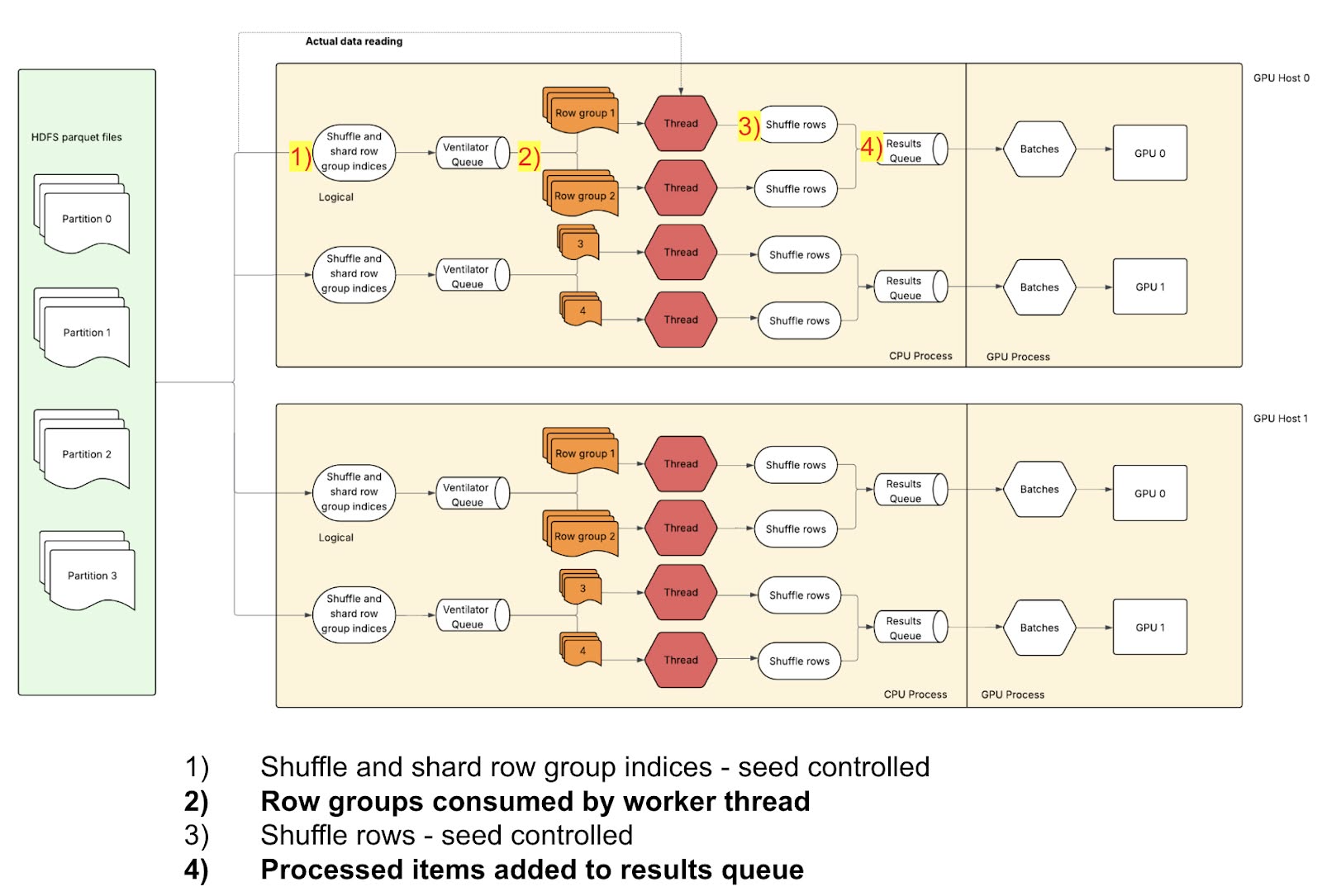}}
\caption{Baseline Petastorm architecture utilizing a shared ventilator and results queue, leading to non-deterministic race conditions based on thread execution speed.}
\label{fig:shared_queues}
\end{figure}

Furthermore, we identified subtle inconsistencies arising from the system's reliance on the legacy \texttt{np.random.RandomState} API, which can behave unpredictably when reseeded across distributed computational components.

\subsection{Deterministic Scheduling Architecture}
To eliminate these sources of randomness without degrading the throughput gains achieved in Section III, we implemented a two-pronged architectural overhaul.

First, we modernized the system's random number generation by deprecating legacy APIs in favor of \texttt{np.random.default\_rng}, ensuring uniform and consistent shuffle operations across the distributed cluster under fixed seeds.

Second, and most critically, we redesigned the worker scheduling topology. We eliminated the shared queues entirely, instead instantiating dedicated ventilator and result queues for each individual worker thread. Work (row groups) is now assigned using a strict, fixed round-robin strategy. Once processed, the transformed results are merged back together for GPU ingestion utilizing the exact same deterministic round-robin order.

\begin{figure}[htbp]
\centerline{\includegraphics[width=\columnwidth]{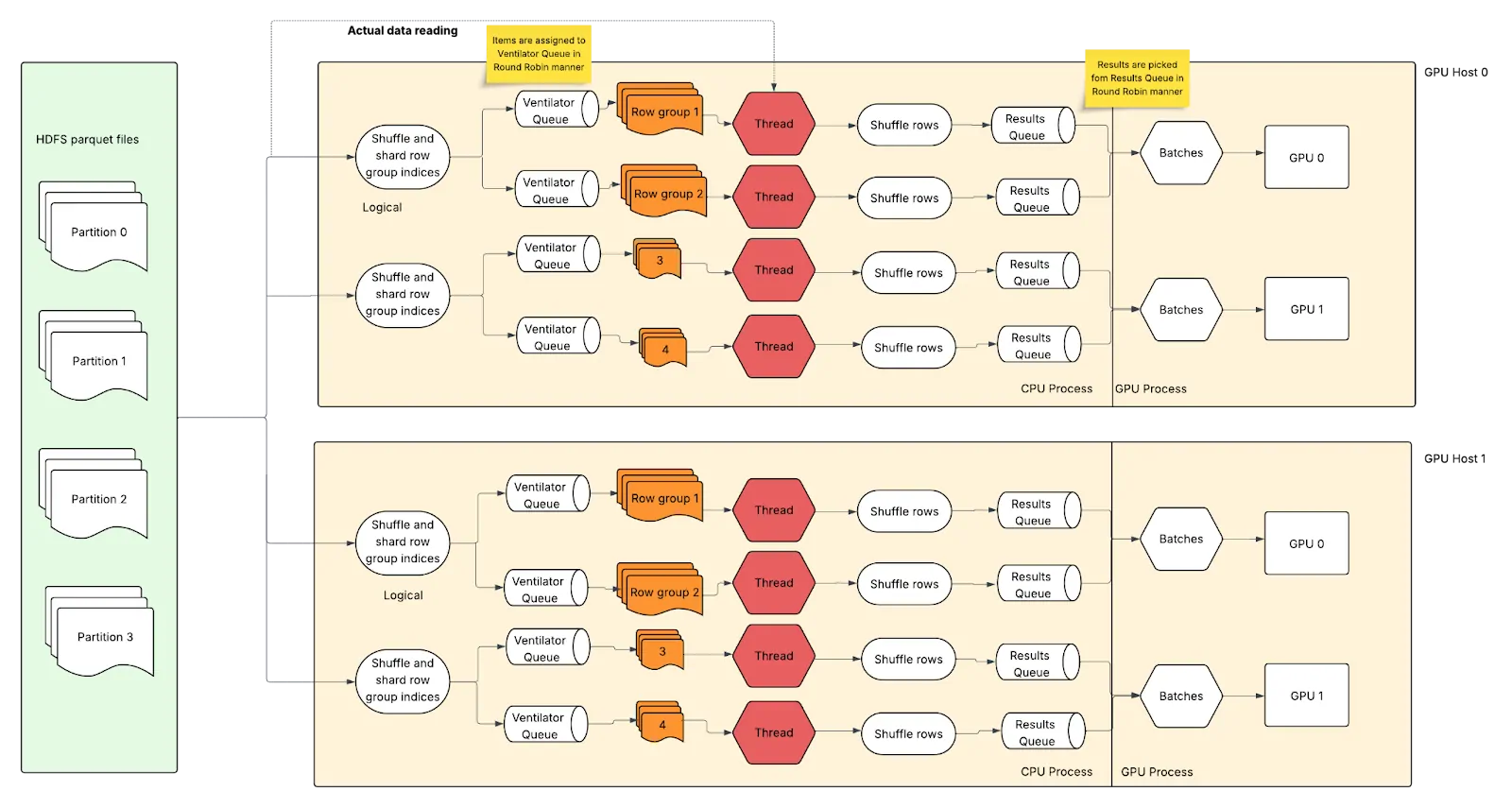}}
\caption{Optimized architecture featuring dedicated queues and round-robin scheduling, guaranteeing strict data determinism without sacrificing parallel throughput.}
\label{fig:dedicated_queues}
\end{figure}

By decoupling the final batch order from individual worker execution speeds, this architecture entirely neutralizes race conditions. Data loading is now strictly deterministic, ensuring that multi-worker training jobs with identical configurations will observe the exact same sequence of training data, independent of hardware or network fluctuations.

\section{Empirical Evaluation}

\subsection{Experimental Setup}
To evaluate the efficacy of our proposed architecture, experiments were conducted using large-scale deep learning recommendation models. The training corpus comprised massive tabular datasets stored in HDFS Parquet format, containing tens of billions of rows and hundreds of features, with total sizes spanning tens of terabytes. The infrastructure utilized multi-GPU compute instances orchestrated via a Ray cluster, with distributed gradient synchronization managed by Horovod.

\begin{table}[htbp]
\caption{System \& Workload Specifications}
\begin{center}
\begin{tabular}{|l|l|}
\hline
\textbf{Component} & \textbf{Specification} \\
\hline
\hline
Dataset Size & Tens of Terabytes \\
\hline
Dataset Scale & Tens of billions of rows, hundreds of features \\
\hline
Storage Format & Apache Parquet\texttrademark{} on HDFS \\
\hline
Compute Hardware & Multi-GPU Instances (High-Performance GPUs) \\
\hline
Cluster Manager & Ray\texttrademark{} \\
\hline
Distributed Training & Horovod \\
\hline
Data Loader & Petastorm (Optimized) \\
\hline
Baseline Epoch Time & $\sim$22 Hours \\
\hline
Optimized Epoch Time & $\sim$3 Hours \\
\hline
\end{tabular}
\label{tab:system_specs}
\end{center}
\end{table}

\subsection{Throughput and GPU Utilization}
The primary objective of Challenge I was to saturate the compute hardware by alleviating I/O and CPU transformation bottlenecks. 

\begin{figure}[htbp]
\centerline{\includegraphics[width=\columnwidth]{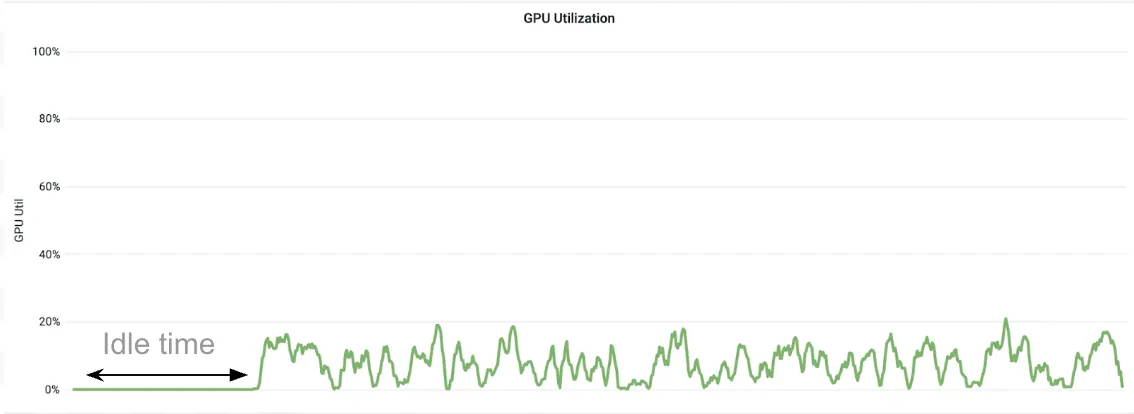}}
\caption{Baseline execution performance. GPU utilization averages 12\% over a 22-hour training period due to data pipeline starvation.}
\label{fig:utilization_baseline}
\end{figure}

Under the baseline architecture, the training pipeline severely throttled compute resources. As shown in Figure~\ref{fig:utilization_baseline}, GPU utilization hovered at an average of 12\%, resulting in an extended end-to-end training time of approximately 22 hours per epoch. 

\begin{figure}[htbp]
\centerline{\includegraphics[width=\columnwidth]{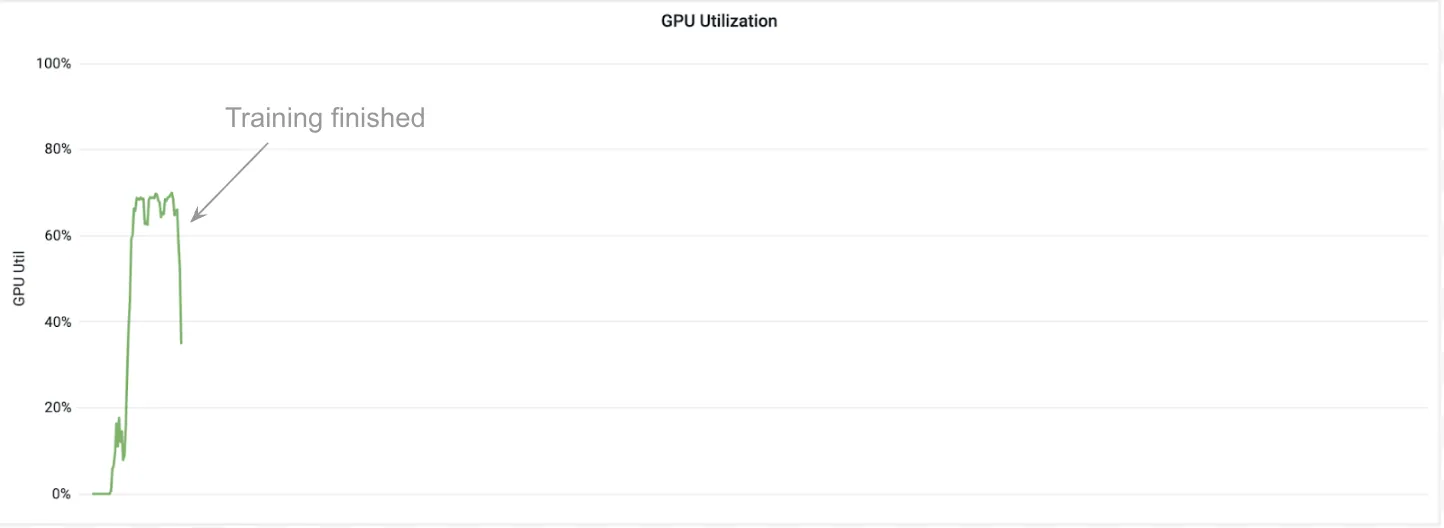}}
\caption{Optimized execution performance. GPU utilization increases to over 60\%, reducing total training time to approximately 3 hours.}
\label{fig:utilization_optimized}
\end{figure}

Following the implementation of push-down transformations and quota-managed local disk caching, system performance improved dramatically. As illustrated in Figure~\ref{fig:utilization_optimized}, average GPU utilization surged to over 60\%, representing a 5$\times$ to 6$\times$ increase in hardware efficiency. 

This improved throughput reduced the end-to-end training time from 22 hours to just 3 hours (an overall 6$\times$ speedup). Beyond significantly accelerating the developer iteration cycle, saturating the GPUs and drastically reducing instance runtimes slashed overall compute costs by nearly 80\%.

\subsection{Reproducibility Analysis}
The primary objective of Challenge II was to guarantee deterministic data loading to enable reliable offline model evaluation.

\begin{figure}[htbp]
\centerline{\includegraphics[width=\columnwidth]{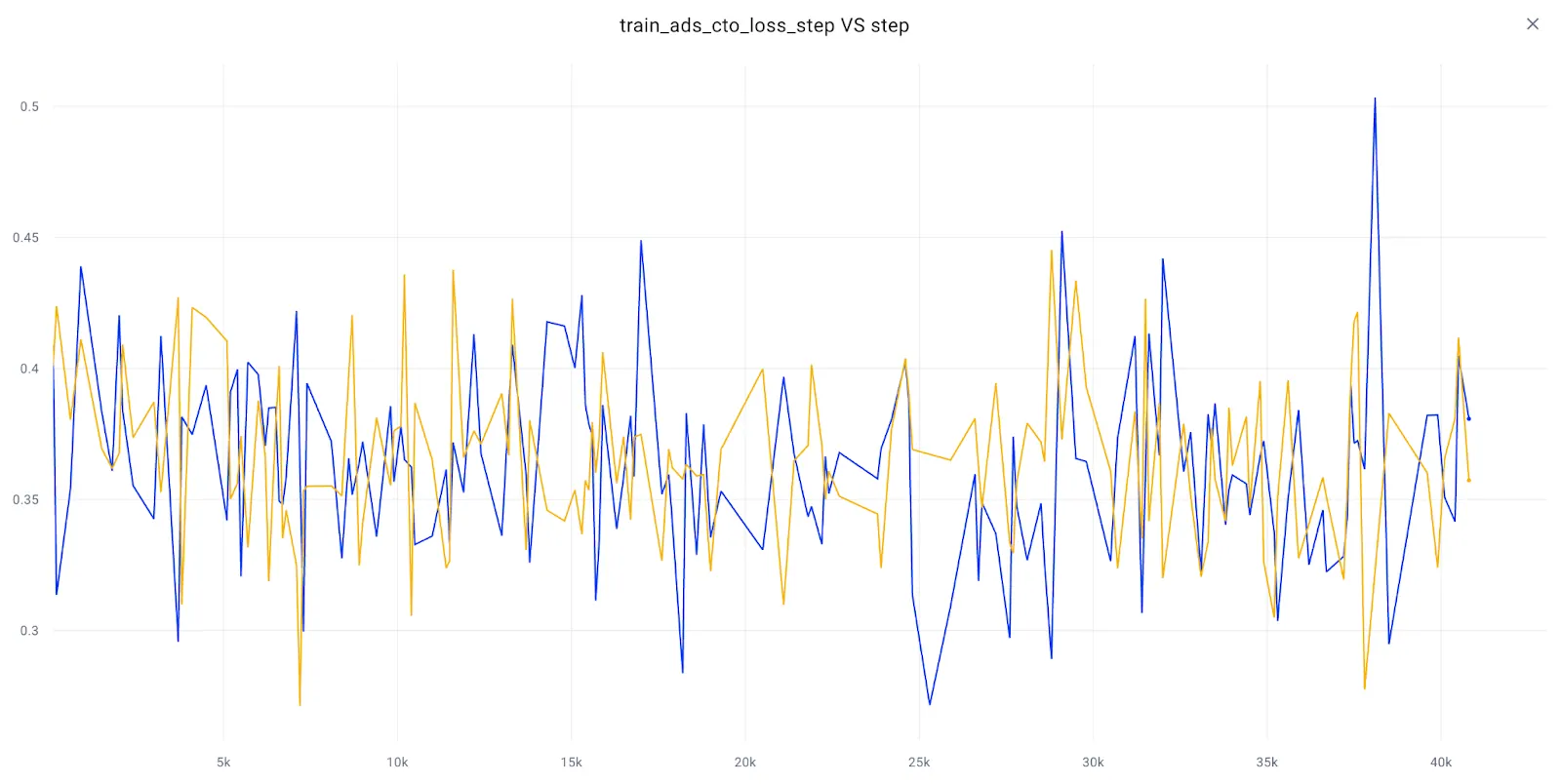}}
\caption{Training loss versus step under the baseline architecture, demonstrating high variance between runs and a MAP shift of $\sim$0.5\%.}
\label{fig:loss_baseline}
\end{figure}

In the baseline state, the race conditions in the shared multi-worker queues caused identical training configurations to process data batches in highly variable sequences. This stochasticity manifested as significant variance in training loss trajectories (Figure~\ref{fig:loss_baseline}), leading to a run-to-run mean average precision (MAP) shift of approximately 0.5\%. At industrial scales, a 0.5\% variance effectively masks the true performance gains of novel architectural changes.

\begin{figure}[htbp]
\centerline{\includegraphics[width=\columnwidth]{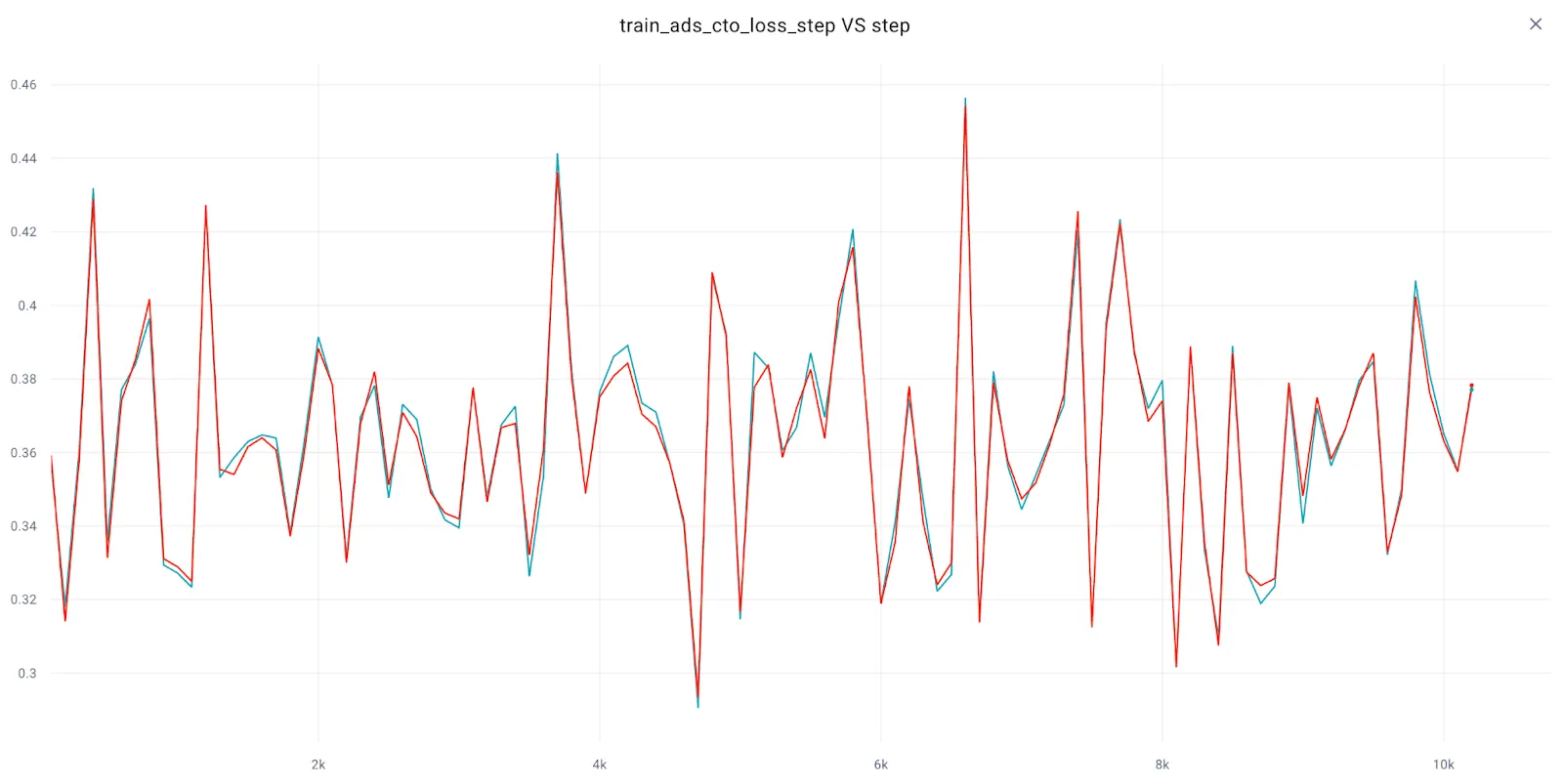}}
\caption{Training loss versus step under the optimized deterministic architecture, showing highly reproducible runs and a reduced MAP shift of $\sim$0.13\%.}
\label{fig:loss_optimized}
\end{figure}

By enforcing strict deterministic scheduling via dedicated round-robin worker queues and modernized RNG initialization, we eliminated dataloader-induced randomness. As shown in Figure~\ref{fig:loss_optimized}, the training loss curves between independent runs are now highly aligned. 

Combined with auxiliary training stability techniques, this dataloader determinism successfully reduced the run-to-run MAP shift from $\sim$0.5\% to $\sim$0.13\%, restoring our ability to confidently conduct A/B testing on model architectures.

\section{Future Work}
While the optimizations presented in this paper have significantly accelerated our internal training pipelines, several avenues for future work remain. Our immediate focus is on ensuring a seamless, platform-wide rollout, establishing this optimized architecture as the default data-loading behavior for all deep learning models across the organization. 

Furthermore, because Petastorm is an open-source framework widely utilized across the industry, we are actively preparing to open-source these architectural improvements. We plan to collaborate with the broader engineering community to integrate these deterministic, high-throughput capabilities into the public Petastorm ecosystem, extending these benefits to external organizations facing similar infrastructure bottlenecks.

\section{Conclusion}
At industrial scale, machine learning efficiency requires a holistic understanding of the interactions between hardware compute, distributed software frameworks, and underlying data formats. This work underscores the critical importance of full-stack profiling; we demonstrated that some of the most substantial gains in training throughput stem not from novel model architectures, but from optimizing the data delivery pipelines that feed them. 

By introducing push-down transformations and quota-managed local caching, we successfully dismantled profound I/O and CPU bottlenecks, achieving a 6$\times$ speedup and a nearly 80\% reduction in compute costs. Concurrently, by redesigning the multi-worker scheduling architecture with dedicated queues and deterministic round-robin assignment, we eliminated hidden concurrency race conditions. This strictly enforced dataloader determinism and restored crucial reproducibility for offline model evaluation.

\section*{Acknowledgment}
The authors would like to thank the Matching ML and Michelangelo leadership teams---specifically Mas-ud Hussain, Dorna Bandari, and Zhitao Li---for their guidance, support, and resource allocation in prioritizing these infrastructure improvements. Their sponsorship was instrumental in transitioning a targeted engineering optimization into a generalized, platform-wide enhancement.

\end{document}